\documentclass[manuscript]{aastex}
\usepackage{graphicx}
\usepackage{color}
\begin{document}

\shorttitle{Fermi Bubble as a source of CRs for $E>10^{15}$ eV}
\shortauthors{Cheng et al.}

\title{THE FERMI BUBBLE AS A SOURCE OF COSMIC RAYS IN THE ENERGY RANGE $>10^{15}$ eV}
\author{K.-S. Cheng$^{1}$, D. O. Chernyshov$^{1,2,3}$, V. A. Dogiel$^{1,2,3}$,
C.-M. Ko$^{1,3}$,
W.-H. Ip$^{3}$, and Y. Wang$^{1}$}
\affil{$^1$Department of Physics, University of Hong Kong, Pokfulam Road, Hong Kong, China}
\affil{$^2$I. E. Tamm Theoretical Physics Division of P.N.Lebedev Institute of Physics,
Leninskii pr. 53, 119991 Moscow, Russia}
\affil{$^3$Institute of Astronomy, National Central University, Jhongli 320, Taiwan; \\
cmko@astro.ncu.edu.tw}


\begin{abstract}

The {\it Fermi} Large Area Telescope has recently discovered two giant gamma-ray bubbles
which extend north and south of the Galactic center with diameters and heights of the order of
$H\sim 10$ kpc.
We suggest that the periodic star capture processes by the Galactic supermassive black hole
Sgr A$^*$, with a capture rate of $\tau_{\rm cap}^{-1}\sim 3\times 10^{-5}$ yr$^{-1}$ and an
energy release of $W\sim 3\times 10^{52}$ erg per capture, can result in hot plasma
injecting into the Galactic halo at a wind velocity of $u\sim 10^8$ cm s$^{-1}$.
The periodic injection of hot plasma can produce a series of shocks.
Energetic protons in the bubble are re-accelerated when they interact with these shocks.
We show that for energy larger than $E> 10^{15}$ eV,
the acceleration process can be
better described by the stochastic second-order Fermi acceleration.

We propose that hadronic cosmic rays (CRs) within the ``knee'' of the observed CR spectrum are
produced by Galactic supernova remnants distributed in the Galactic disk. Re-acceleration
of these particles in the Fermi Bubble produces CRs beyond the knee.
With a mean CR diffusion coefficient in this energy range in the bubble
$D_B\sim 3\times 10^{30}$ cm$^2$ s$^{-1}$, we can reproduce the spectral index of the spectrum
beyond the knee and within.
The conversion efficiency from shock energy of the bubble into CR energy is
about 10\%.
This model provides a natural explanation of the observed CR flux, spectral indices,
and matching of spectra at the knee.

\end{abstract}



\date{\today}

\maketitle

\section{INTRODUCTION}\label{sect:intro}

The recent discovery of a pair of giant Fermi Bubbles in the Galactic center (GC)
is one of the most remarkable events in astrophysics.
The first indication of the structure appeared in a paper by \citet{Dobler2010},
which they called the Fermi Haze.
Using the special procedure of background subtraction of the {\it Fermi}
Large Area Telescope (LAT) data, \cite{meng}
discovered a pair of symmetric structures above and below the Galactic plane
in the GC direction.
The origin of the bubble,
if its existence can be proved, is still enigmatic, and up to now a few models have been
presented in the literature.
The team that subtracted this structured gamma-ray emission from the total diffuse Galactic
emission presented different explanations of the phenomenon, but they seemed to favor the model
that a single massive release of energy in the GC when a huge cloud of gas or a star cluster was
captured by the central supermassive black hole about 10 Myr ago.

A similar explanation was suggested by \citet{guo11}
and \citet{guo2011}.
They assumed that the Fermi Bubbles were created by a recent active galactic nuclei
jet activity about 1-2 Myr ago, which was active for a duration
of $\sim$ 0.1-0.5 Myr, releasing energy totaling $\sim$ (1-8)$\times 10^{57}$ erg.
The bipolar jets were ejected into the Galactic halo
along the symmetric axis perpendicular to the Galactic plane.

It is important to note that the existence of the bubbles was first evidenced in
X-rays detected by {\it ROSAT} as a narrow envelope with very sharp edges
\citep{bland2003} and later the {\it Wilkinson Microwave Anisotropy Probe}
({\it WMAP}) detected an excess of radio signals at the location of the
gamma-ray bubbles
\citep{fink,Dobler2008,Dobler2011}.
The {\it ROSAT} structure is
explained as a fast wind with a velocity $u_{w} \sim 10^8$ cm s$^{-1}$
driving a shock into the halo gas. This phenomenon
requires an energy release of about $10^{55}$ erg at the GC and
this activity should be periodic on a timescale of the order of 10 Myr.
This requirement of energy release in the GC is consistent with the observations of the
existence of hot plasma with a temperature of about 10 keV in the
GC region but with a radius of 30-50 pc only.
This cannot be confined and will escape
from the GC with a speed of $u_{w} \sim 10^8$ cm s$^{-1}$ \citep{koyama07}.
Therefore, sources with a power of about $10^{41}$ erg s$^{-1}$ must have heated the
plasma or released $\sim 10^{55}$ erg from
the GC in the past 10 Myr. However, \citet{crock} have proposed a
relatively slow energy release ($\sim 10^{39}$ erg s$^{-1}$) from
supernova (SN) explosions as a source of proton production in the GC.
The observed gamma rays come from hadronic processes of the protons in the halo.
The plasma in the halo is extremely turbulent and the protons are trapped for a time
comparable to the Hubble time.
But this model requires a separate origin of electrons, which have a much shorter
lifetime than protons, to explain the {\it WMAP} data.

Intensive energy release has been observed, indeed, at the
center of normal galaxies as strong variations of X-ray
radiation.  There are common characteristics of these X-ray
sources. First, all of them have been bright sources, and
their X-ray luminosity could go up to about $10^{44}$ erg s$^{-1}$.
Second, they have shown a high level of variability in their X-ray
light curve within years. In the ``high state'', the luminosity of
one source could be at least 100 times higher than the luminosity
in its ``low state''. Third, most of them have a super soft
spectrum during the flare, with effective blackbody temperatures
of only about 10-100 eV \citep{Komossa,Halpern}.
The classic examples that satisfy these characteristics are
RX J1624.9+7554, RX J1242.6-1119A, RX J1420+5334, RX J1331-3243,
and NGC 5905. Many scenarios have been proposed to explain these
phenomena, but most of them fail to explain some of the
observed results. A detailed discussion of these scenarios can be
found in \citet{Komossa}. Among all of the listed models, the
tidal disruption model is the most commonly accepted, and it
gives the most satisfactory explanation to the observations by
considering the radiation from the disk. In this model, when a
star passes by a black hole within a capture radius, where the
black hole tidal force becomes stronger than the self-gravity of
the star, the star can be captured. The detailed capture and
disruption process of a main-sequence star has been studied by
several authors \citep[e.g.,][]{Rees1988,Cannizzo}.
The capture rate of main-sequence stars in our Galaxy and in other galaxies
is about $10^{-4}$ yr$^{-1}$ to $10^{-5}$ yr$^{-1}$ \citep[see][]{syer1999,alex05}.
Recently more stellar capture events have been observed
\citep[e.g.,][]{Esquej,Gezari,Gezari2009,Komossa,Cappelluti}.
Dynamical studies of nearby galaxies
suggest that most, if not all, galaxies with a bulge component host
a central supermassive black hole and that the bulge and black
hole mass are tightly correlated \citep{Magorrian,Tremaine,Greene}.

The Burst Alert Telescope on board {\it Swift} has identified a transient
X-ray source called GRB110328A \citep{Cummings} with later
optical identification \citep{Cenko,Leloudas},
which is located in the direction of the constellation Draco
when it erupted in a series of X-ray flares. The distance to this
source is determined to be $z\sim 0.35$ by using H$_{\beta}$ and
O$\,{\scriptstyle\rm III}$ emission lines by the Gemini telescope \citep{Levan2011b}. The
characteristics of GRB 110328A appear inconsistent with those of a
gamma-ray burst \citep{Barthelmy}. In fact, its time-dependent
characteristics including various timescales in light curve,
multi-wavelengths, etc., seem to be better explained in terms of the
tidal disruption of a star by a supermassive black hole
\citep{Almeida,Bloom,Levan2011a,Burrows,Zauderer}.
All these recent observations suggest that
stellar capture processes are quite common in other normal galaxies.

Observations have also revealed much evidence of unusual
processes occurring in the central region of our Galaxy,
for instance, the enigmatic 511 keV annihilation emission discovered
by the {\it International Gamma-Ray Astrophysics Laboratory}
\citep[see, e.g.,][]{knoed} whose origin is still debated.
The hot plasma has a temperature of about 10 keV which cannot be confined in the GC
and, therefore, sources with a power of about $10^{41}$ erg s$^{-1}$
are required to heat the plasma \citep[see][ and references
therein]{koyama07}. In fact, plasma outflows with velocities of $\ga 10^7$ cm s$^{-1}$
are observed from the nuclear region of our Galaxy
\citep[see][]{crocker2} and from the nucleus of Andromeda \citep{gilf}.
Time variations of the 6.4 keV line and X-ray
continuum emission observed in the direction of molecular clouds
in the GC which are supposed to be a reflection of a giant X-ray flare occurred
several hundred years ago \citep[see][ and references therein]{inui,ponti,terr}.
HESS observations of the GC in the TeV energy range indicated
an explosive injection of cosmic rays (CRs) there, which might be associated with
the supermassive black hole Sgr A$^\ast$ \citep[e.g.,][]{ahar}.

In a series of papers
\citep{cheng1,cheng2,dog_pasj1,dog_pasj,dog_aa,dog_pasj2011}, we
developed a model of energy release in the GC due to
star accretion onto the central black hole for the interpretation
of X-ray and gamma-ray emission from the GC. Our goal was to
explain these observational data in the framework of a single
model. Basic assumptions in these models are (1) the Galactic
supermassive black hole Sgr A$^*$ can capture a star at a rate of
$\nu_s\sim 10^{-4}$-$10^{-5}$ yr$^{-1}$, and (2) the energy release from each capture
in the form of a flux of subrelativistic protons is
$W\sim 4\times 10^{52}\,M_\ast^2R_\ast^{-1} M_{\rm bh}^{1/3}(b/0.1)^{-2}$ erg,
where $M_\ast$ (in units of $M_\odot$) and $R_\ast$ (in units of $R_\odot$)
are the mass and the radius of the captured star,
$M_{\rm bh}$ (in units of $10^6 M_\odot$) is the mass of the supermassive black hole, and
$b$ is the ratio of the periapse distance $r_p$ to the tidal radius $R_T$
\citep[see the review of][]{alex05}. In a time scale much
longer than the capture timescale, this model can be treated to have
an average power injection $\dot{W}\sim 3 \times 10^{40}$ erg s$^{-1}$.
These protons heat the surrounding plasma by Coulomb losses to 10 KeV.

Based on this model, \citet[CCDKI model]{cheng}
argued that up to several hundred capture events might have occured in the past 10 Myr,
which may have generated a series of shocks propagating through the central
part of the Galactic halo and thus produced accelerated
relativistic electrons responsible for the bubble emission.
Processes of charged particle acceleration by the bubble shocks in
terms of sizes of the envelope, maximum energy of accelerated
particles, etc., may differ significantly from those obtained for
SNe. In this paper, we examine whether a ``signal'' from charged particles
accelerated in the bubble region can be seen in the spectrum of
CRs observed at the Earth.  We present simple
estimations of hadronic CR acceleration by the Fermi Bubble shocks up to energies
above $10^{15}$ eV. The paper is organized as follows. In
Section~\ref{sect:snr}, we review current understanding of CR acceleration
by supernova remnants (SNRs) and conclude that this process can only produce CRs with
energies less than $10^{15}$ eV. We present a simple solution of the
multiple-shock structure in the halo in Section~\ref{sect:structure}.
In Section~\ref{sect:bubbleshock} we discuss the protons accelerated by the bubble shocks.
We emphasize that a broken power law of particle distribution must be formed
because of the finite spacing between consecutive shocks and
the spectral break naturally occurs at
$10^{15}(u/10^8\,{\rm cm\, s}^{-1})(l_{\rm sh}/30\,{\rm pc})(B/5\,\mu{\rm G})$ eV.
Charged particles below
and above this critical energy are accelerated by two different
acceleration mechanisms. In Section~\ref{sect:beyondknee}, we calculate the total
particle spectrum by summing up the contribution from all
shocks in the bubble and compare it with the observed hadronic CR
spectrum with energies larger than $10^{15}$ eV.
In Section~\ref{sect:models}, we suggest a model that can produce the CR spectrum
within and beyond the ``knee'' (around $3\times 10^{15}$ eV).
Summary and discussion is presented in Section~\ref{sect:summary}.

\section{CR ACCELERATION BY SNRs IN THE GALAXY}\label{sect:snr}

From a general point of view, SN explosions are enough to supply the
power needed for the luminosity of CRs in our Galaxy,
$L_{\rm CR}\sim 10^{41}$ erg s$^{-1}$
\citep[for a general review see][]{berezh1,Reynolds}.
Diffusive shock acceleration \citep[see][]{krym, bell78} is considered
to be a viable and natural mechanism for CR accelerated by SNRs.
The mechanism produces a power-law spectrum with the necessary spectral index
that is observed experimentally.
The simplest one-dimensional kinetic equation describing this process has the form
\begin{equation}
\frac{\partial f}{\partial t}+\frac{\partial}{\partial z}
\left(u(z)f-D\frac{\partial f}{\partial z}\right)
-\frac{1}{3p^2}\frac{du(z)}{dz}\frac{\partial}{\partial p}\left(p^3f\right)=0\,, \label{shock}
\end{equation}
where $z$ is the coordinate perpendicular to the shock front, $p$
is the particle momentum, $u(z)$ is the velocity distribution
which describes a velocity jump at the shock, and $D$ is the
spatial diffusion coefficient. The solution of this equation is a
power-law function, $f(p)\propto p^{-\gamma}$, in which the spectral index
$\gamma$ is a function of the velocity jump at the shock.
For strong shocks with a Mach number much larger than unity, $\gamma=4$.
The corresponding energy spectral index is $\nu=2$ ($N(E)\propto E^{-\nu}$).

The current status of the observations of middle-aged SNRs by LAT on broad {\it Fermi}
with an energy range from $0.2$ to $100$ GeV has provided some insight into the shock-acceleration
theory of SNRs \citep{Castro,Uchiyama}.
Assuming the gamma rays are produced by hadronic processes,
\citet{Castro} deduced the spectral index of four SNRs ranging between $2.1$ and $2.4$.
However, whether the observed GeV gamma rays are produced by hadronic processes or
leptonic processes is very difficult to differentiate.
On the other hand the ambiguity can be removed if broadband emissions are observed.
In particular if GeV and TeV spectra can be described by a single power law,
which is steeper than $E^{-2}$, the hadronic processes could be the more favorable mechanisms.
Currently about 10 SNRs have been detected in both GeV and TeV bands,
including Tycho and CTB37A, whose GeV-TeV gamma-ray emission shows uniformly
steep spectral indices of about 2.3 and 2.2, respectively
\citep[see Table 1 of ][]{Damiano}.
All these recent observations are consistent with conventional SN shock-acceleration
theories, which suggest that the spectrum of CRs is roughly described by $E^{-2}$.

Many fundamental questions related to the assumption that SNRs are
the sources of Galactic CRs are still open. The maximum
energy of the accelerated particles is the main concern for this
scenario, which can be roughly estimated from a very simple
relation. The acceleration time at the shock is $\tau_{\rm acc}(E)
\sim D(E)/u^2_ {\rm sh}$, where $u_{\rm sh}$ is the shock velocity ($\sim$
a few $10^8$ cm s$^{-1}$). The minimum value of the diffusion coefficient
at the shock follows from the Bohm diffusion scenario, i.e.,
$D_{\rm Bohm}(E) = (c/3)r_L(E)$ where $r_L$ is the Larmor radius of the
particle. Equating the acceleration time with the lifetime of the shock $T$
we get an estimate for the maximum energy of accelerated particles after time $T$,
\begin{equation}\label{emax}
E_{\rm max} \sim Ze\beta_{\rm sh}u_{\rm sh}  B  T \,,
\end{equation}
where $\beta_{\rm sh}=u_{\rm sh}/c$, $B$ is the magnetic field strength at the shock.
The combination $u_{\rm sh} B/c$
can be interpreted as an effective electric field.

For an SNR of typical age $\tau_{\rm SNR}\sim 1000$ yr,
the maximum energy of protons is easily estimated by
requiring that the acceleration time remain smaller than
$\tau_{\rm SNR}$. \citet{laga} and \citet{volk} demonstrated that the
maximum energy of protons within the scenario of Bohm diffusion is
as large as $E_{\rm max} \sim 10^{13}$-$10^{14}$ eV for standard
galactic SNRs. \citet{berezh1} estimated the efficiency of
acceleration when a feedback reaction of accelerated particles on
the front structure was included and they showed that in the Bohm
limit CRs absorb about 20\% of the explosion energy. The
acceleration process acts as an effective viscosity in widening the
region of the shock velocity jump
and eventually the acceleration process stops.

However, outside the quasi-linear model the acceleration of CRs
at the shock fronts of SNRs may make the acceleration of
particles more effective. As \citet{bell} \citep[see also][]{byk09}
showed that during acceleration at shocks of SNR, the
magnetic non-resonant fluctuations were strongly driven. A
nonlinear MHD simulation indicated that CR-excited turbulence
could amplify the magnetic field. It appears that acceleration to
the spectral break at $10^{15}$ eV normally ceases as an SNR enters
the Sedov phase. Thus, CR acceleration by SN shocks can only
provide particles with energies less than $10^{15}$ eV.

The spectral index of the observed CR flux changes from 2.7 to 3.1
around energy $10^{15}$ eV, and this is known as the knee.
The standard model of CR acceleration by SN shocks cannot explain
CR energies above the knee, because it only produces a single power-law
spectrum up to the energy around $10^{15}$ eV.
In addition, the CR spectrum flattens again for energies above $10^{18}$ eV,
and this is known as the ankle.
Large size is required to accelerate and
to confine charged particles above the ankle (the Larmor
radius at these energies is comparable with the halo height).
The origin of CRs above the ankle is generally attributed to
an extragalactic origin because those particles could not be confined
inside the Galaxy and known potential galactic accelerators
could hardly accelerate particles to such high energies.

The origin of the steepening for $E\ga 10^{15}$ eV is still an
open question and different mechanisms of CR acceleration in the
range $10^{15}$-$10^{18}$ eV in the Galaxy have been proposed.
\citet{ptus10} assumed that the CR flux with energies above
$10^{15}$ eV is produced by very young galactic SNRs. They modeled the
particle acceleration by spherical shocks with back-reaction
of CR pressure on the shock structure. The significant
magnetic field amplification in young SNRs produced by CR
streaming instability may lead to a flux of CRs with the maximum
energy of accelerated particles about $5\times 10^{18}$ eV. In this
model, the steepening of the CR spectrum at the knee position is
due to distortion of the spectrum ejected from young SNRs by the
propagation process.

Another interpretation was suggested by \citet{wolf1}, \citet{wolf2},
and \citet{lagut}
who assumed that CRs at the knee were produced by a single, recent
local SN.
Recently \citet{butt} summarized problems (including energies within the knee)
of the conception that isolated SNRs are the main accelerators of CRs and
discussed alternative scenarios of CR acceleration.

The steepening of the CR spectrum at high energies may
also be the result of a change in properties of diffusion in the
interstellar medium. This effect of propagation was mentioned
first by \citet{syr71} who noticed that the standard diffusion of
CRs in the interstellar medium might be changed by convection  due
to a drift of these particles in the large-scale Galactic magnetic
fields. This model was developed in \citet{ptus93} who assumed two
types of CR diffusion in the Galaxy: the usual diffusion due to
particle scattering on fluctuations of random magnetic fields and
the Hall anisotropic diffusion (drift motion) due to the large-scale
Galactic magnetic field whose effect might become important
just above the knee energy.

As an alternative model, \citet{joki} suggested a mechanism of acceleration
in the Galaxy of ultra high energy CRs in a Galactic wind
and its hypothetical termination shock. In this scenario, SNRs
accelerate the bulk of CRs up to $10^{15}$ eV. These
particles are further accelerated up to $10^{19}$-$10^{20}$ eV at a
termination shock which is at a distance of a few hundred kpc
from the disk. \citet{ip92} analyzed multiple
interactions of particles with SNRs in the Galactic disk as a
source for CR acceleration above the knee.
However, too many shocks are required in the disk in order to produce CR flux at the knee.
\citet{byk90} showed that regions of CR acceleration
to energies above $10^{15}$ eV might be OB associations where concentration
of shock fronts is very high.
We will discuss this model in Section~\ref{sect:bubbleshock}.

In summary, it is generally agreed that SNR shocks can only accelerate particles to
energies less than $10^{15}$ eV.
On the other hand, accretion processes in the GC may generate giant shocks which are
effective for particle acceleration above $10^{15}$ eV.
It is interesting to note that the notion of recurrent activities in the Galactic center as
a source of cosmic rays has been proposed some 30 years ago
\citep{Khazan1977,Said1981,Ptuskin1981},
although not in the same perspective as the present paper.

\section{STRUCTURE OF SHOCKS IN THE FERMI BUBBLE}\label{sect:structure}

As was assumed in CCDKI the central massive black hole captures
a star every $\tau_{\rm cap}\sim 3\times 10^4$ yr; as a result,
about $W \sim 3 \times 10^{52}$ erg of energy in the form of subrelativistic particles
is released.
This heats up the surrounding gas in the central region of our Galaxy.
The hot gas expands into the halo and forms a propagating upward shock.
The situation is very similar to that of the stellar wind of a massive star blowing into its
surrounding medium \citep[see, e.g.,][]{weav77,kogan}.

The gas distribution in the disk and in the halo was derived in
\citet{cordes} and can be presented as a double exponential
distribution
\begin{eqnarray}\label{disk}
n(\rho,z) = &&0.025 \exp\left(- \frac{z}{1\,{\rm kpc}}\right) \times
\exp\left[- \left(\frac{\rho}{20\,{\rm kpc}}\right)^2\right] \nonumber\\
&& +\,0.2 \exp\left(- \frac{z}{0.15\,{\rm kpc}}\right)\times
\exp\left[-\left( \frac{\rho-4\,{\rm kpc}}{2\,{\rm kpc}}\right)^2\right]
\ {\rm cm}^{-3}\,.
\end{eqnarray}

The energy release in the GC as a result of star capture can be either
impulsive or continuous depending on the characteristic times of
star capture and energy dissipation of subrelativistic protons
(plasma heating by Coulomb losses). The capture time is roughly
$\tau_{\rm cap}\sim 3 \times 10^4$ yr and the dissipation time $\tau_{\rm diss}$ in the CCDKI
model is determined by the rate of ionization losses of protons
injected with energy $E_p$, which is given by
\begin{equation}
\tau_{\rm diss}\simeq 10^6
\left(\frac{n}{1\,{\rm cm}^{-3}}\right)^{-1}\sqrt{\frac{E_p}{100\,{\rm MeV}}}\ {\rm yr}\,,
\end{equation}
where $n$ is the gas density in the vicinity of the GC which can be quite high
\citep[see the discussion in][]{cheng2}, e.g., in nearby molecular clouds $n>10^4$ cm$^{-3}$.

If $\tau_{\rm cap}\ll\tau_{\rm diss}$ we have the case of stationary energy
injection in the GC. In this case the region of heated gas is
bounded by a single shock \citep[see][]{weav77}.
For $\tau_{\rm cap}\gg\tau_{\rm diss}$ a multi-shock structure is formed in the
halo with  shocks of different ages.
A similar multi-shock structure can also be created if there are epochs of
high-frequency star captures in the GC.
Thus, the number of shocks is determined by the injection and dissipation parameters.

\begin{figure}
\epsscale{.80}
\plotone{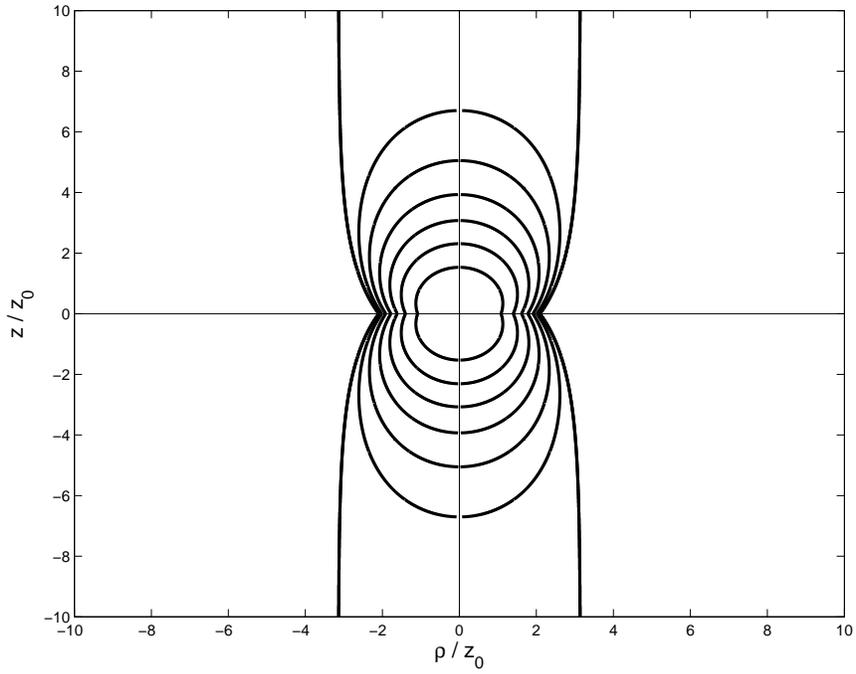}
\caption{
Multi-shock structure in the bubble from the Kompaneets' solution showing
seven representative shocks.
}
\label{fig:shape}
\end{figure}

For a highly simplified case of the exponential atmosphere with the scale height $z_0$,
i.e., the plasma density $n(z)= n_0 \exp(-z/z_0)$,
an  analytic  solution of shock propagation
was obtained by \citet{komp} \citep[see also the review of][]{kogan}.
This solution gives a qualitative picture of the shock propagation and parameters of the
medium bounded by the shock that roughly described the situation expected in the Fermi Bubble.
If the rate of energy injection is  $L$ then the radius of the shock as a function of
the height $z$ and the time $t$ is
\begin{equation}
\rho(z,t)=2z_0\arccos\left\{\frac{1}{2}\exp\left(\frac{z}{2z_0}\right)
\left[1-\left(\frac{y}{2z_0}\right)^2+
\exp\left(-\frac{z}{z_0}\right)\right]\right\}\,. \label{rz}
\end{equation}
Here $z$ is the coordinate perpendicular to the Galactic plane,
\begin{equation}
y=\int\limits^t_0\left[\frac{(\gamma_p^2-1)}{2}\frac{\lambda
Lt}{V(t)mn_0}\right]^{1/2}dt\,,
\end{equation}
$V(t)$ is the current volume enveloped by the shock
\begin{equation}
V(t)=2\pi\int\limits^{a(t)}_0\rho^2(z,t)\,dz\,,
\end{equation}
$a(t)$ is the position of the top of the shock
\begin{equation}
a(t)=-2z_0\ln\left(1-\frac{y}{2z_0}\right)\,, \label{at}
\end{equation}
$L=W/\tau_{\rm cap}$ is the average luminosity of the central source,
$\gamma_p$ is the polytropic coefficient, and $\lambda$ describes the fraction
of explosion energy converted into the thermal energy of gas \citep[see][]{kogan}.
As it follows from Equations~(\ref{rz}) and (\ref{at}), for the finite time $t_1$ determined from
the condition $y(t_1)=2z_0$ the shock breaks through the exponential atmosphere and the
bubble top $a(t_1)$ tends to infinity  while the bubble radius in the Galactic plane ($z=0$)
tends asymptotically to $\rho=2z_0\cos^{-1}(1/2)\simeq 2z_0$ and for $z\gg z_0$ to
$\rho\simeq\pi z_0$.

A numerical solution of the system (\ref{rz})-(\ref{at}) in dimensionless coordinates
for shock waves at different ages is shown in Figure~\ref{fig:shape}.
This figure is meant to be illustrative only. In reality the distribution of shocks should be
far more complicated.
Neglecting the thin 4 kpc torus component of Equation~(\ref{disk}), the scale height of the
``atmosphere'' is $z_0=1$ kpc.
The shock distribution in Figure~\ref{fig:shape} is suggestive of a cylindrical bubble with
an edge at a radius $\rho_B\simeq 3$ kpc.

Based on this, we put forward a quasi-stationary model of the Fermi Bubble which we regard
as the source of hadronic CRs with energy larger than $10^{15}$ eV.
The essence of the CCDKI model is that energy is quasi-periodically injected into the
halo when the stellar capture processes take place and may exist over a timescale
comparable with the age of Milky Way.
Consequently, the Fermi Bubble should have a stationary structure.
The idealized Kompaneets' solution above shows that there is a stationary
sideway boundary for shocks.
For quasi-periodic star capture, the bubble interior is filled with shocks propagating in
series and eventually stopping at $\rho_B\simeq 3$ kpc.
However, in the Kompaneets' solution dissipation processes are ignored.
The realistic situation has to be described by a set of dissipative hydrodynamic equations,
which takes account the shock propagation in a non-uniform medium and various
dissipation processes, including shock heating, energy transfer into CRs,
slowing down due to accumulation of material, etc.
Fitting with the observed gamma-ray spectrum, \citet{cheng} concluded that electrons
should have an escape time scale of 15 Myr in order to explain the spectral break position.
The characteristic dissipation time scale of the shocks should be of the same order
if these electrons are transported away by the shocks.
The shocks will be mostly dissipated when they arrive at the sideway boundary
($\rho_B$) and
there will be no pile-up of shocks at $\rho_B$. As the speed of shocks along the bubble axis is
progressively larger than the sideway speed, the upper and lower boundaries of the bubble
will be the same as the halo boundary ($z=\pm H=\pm 10$ kpc from the mid-plane).
Thus, the bubble has a stationary structure.

The same result for the dimension of the bubble can be obtained if we use the swept-up mass model
proposed by \citet{cheng}.
In this model the radius of the sideway shock front or the swept-up front is given by
$\rho_s=\sqrt{2\lambda W/\pi m n \Delta z u_{s\rho}^2}$, where $u_{s\rho}$ is the speed of the
sideway shock front (or the swept-up front) and $\Delta z=u\tau_{\rm cap}$.
With $u\sim 10^8$ cm s$^{-1}$, $\tau_{\rm cap}\sim 3\times 10^4$ yr, $\Delta z\sim 30$ pc.
We argue that the sideway shock front or the swept-up front will disappear when its speed is
smaller than the local sound speed.
In Cheng et al. (2011) we have estimated that the shocks heat up the halo to $\sim 1$ keV
and the characteristic sound speed is of the order of $v_s \sim 3\times 10^7$ cm s$^{-1}$.
Thus putting $u_{s\rho}\sim v_s$, the sideway boundary $\rho_B\approx 3.2\,{\rm kpc}
(\lambda W/2\times 10^{52}\,{\rm erg})^{1/2}
(\Delta z/30\,{\rm pc})^{-1/2}
(n/10^{-3}\,{\rm cm}^{-3})^{-1/2}
(v_s/3\times 10^7\,{\rm cm\,s}^{-1})^{-1}$
(note that $\lambda W$ is the fraction of injected energy converted into thermal energy of gas).

\section{PROTON ACCELERATION BY THE BUBBLE SHOCKS}\label{sect:bubbleshock}

Correct analysis of shock acceleration in the bubble requires
sophisticated calculations in each stage of this process which we
perform latter. Now we present simple estimates of the
characteristics of the spectra of the accelerated particle in the framework of the
CCDKI model.

Below we analyze the spectrum of
protons accelerated in the bubble and discuss whether the bubble's
contribution to the total flux of CRs in the Galaxy may explain
the knee steepening. We remind the reader that the generally accepted point
of view is that the flux of relatively low energy CRs
($<10^{15}$ eV) is generated by SNRs, which eject a power-law
spectrum $E^{-2}$ into the interstellar medium. This spectrum is
steepened by propagation (escape) processes in the Galaxy in
accordance with the spectrum observed near Earth \citep[for
details, see][chap. 5, Section 6]{ber90}. However these sources can hardly produce CRs
with energies $>10^{15}$ eV, at which a
steepening (the knee) in the CR spectrum is observed. For
characteristics of the knee spectrum and models of its origin see
the review of \citet{kumiko}. We suggest that the bubble could
generate the flux of CRs at energies $> 10^{15}$ eV because the shocks in
the bubble have much larger length scales and longer lifetimes in comparison with
those in SNRs.

In the framework of CCDKI, the bubble may fill with hundreds of
shocks propagating in series one after another,
though a single shock structure cannot be excluded.
The average separation between two shocks is given by
\begin{equation}
l_{\rm sh}=\tau_{\rm cap}u=30\,\left({\tau_{\rm cap}\over 3\times 10^4\,{\rm yr}}\right)
\left({u\over 10^8\,{\rm cm\,s}^{-1}}\right)\ {\rm pc}\,.
\label{l_shock}
\end{equation}
However, the exact amount of time between two consecutive shocks
depends on the actual time between two consecutive capture
events and their energy releases. There is another important
spatial scale which characterizes processes of particle
acceleration by a single shock: the diffusion length scale at a single shock
$l_D \sim D/u$. Here $u$ is the shock velocity and $D$ is the spatial diffusion coefficient
of the energetic particles near a shock which depends on particle interaction with
small-scale magnetic fluctuations.
In the Bohm limit, $D\sim c r_L(E)/3$, where $r_L(E)=E/ZeB$ is the
particle Larmor radius. In this case
\begin{equation}
l_D\sim{cr_L\over u}={cE\over ZeB u}\,.
\label{l_acc}
\end{equation}

The problem of particle acceleration in conditions of supersonic
turbulence (multiple-shock structure) has been extensively analyzed
\citep[e.g.,][]{Spruit,Achterberg,Schneider,Melrose} as well as quasi-periodic flows
\citep[e.g.,][]{Webb2003}.
In a series of papers by \citet{byk90}, \citet{byk92} and \citet{byk01}
the idea was applied to acceleration
processes in OB associations, which is quite similar to the structure of the bubble.
They introduced a dimensionless parameter characterizing the acceleration regimes
\begin{equation}
\psi=\frac{l_{\rm sh}}{l_D}\sim\frac{ul_{\rm sh}}{D}\sim \frac{ul_{\rm sh}}{cr_L}\,.
\end{equation}
The critical energy $E_1$ that separate two regimes of acceleration can be
estimated from the condition $\psi\sim 1$ or $l_D (E_1)\sim l_{\rm sh}$.
For the conditions of the Fermi Bubble the critical energy is
\begin{equation}
E_1\approx{ZeBul_{\rm sh}\over c} = 10^{15}\,Z\left({B\over 5\,\mu{\rm G}}\right)
\left({l_{\rm sh}\over 30\,{\rm pc}}\right)
\left({u\over 10^8\,{\rm cm}}\right)\ {\rm eV}\,. \label{e1}
\end{equation}

In the case of $\psi\gg 1$ or $l_D\ll l_{\rm sh}$, the analysis in
\citet{byk90} and \citet{byk92} showed that there is a combined effect of a
fast particle acceleration by a single shock, which generates the
spectrum $E^{-2}$   and relatively slow transformation of this
spectrum due to interaction with other shocks (stochastic Fermi
acceleration) into a hard $E^{-1}$ spectrum in the intershock
medium at relatively low energies. However, it is unclear if such
slow transformation can be completed within the lifetime of the
shocks in the bubble. A detailed numerical analysis is needed.
Furthermore, from the general point of view the characteristic
acceleration time is quite short in the range $E\la E_1$, which is
roughly given by the shock acceleration time $c r_L/u^2$. In the
range $E\ga E_1$, the acceleration time scale increases to the
time of stochastic acceleration, $cl_{\rm sh}/u^2$.
With an average Galactic spatial diffusion coefficient $D_G$ outside the
bubble and a Galactic halo of height $H$, the characteristic escape time
is $\tau_{\rm esc}\sim H^2/D_G$.
We expect that escape processes, which play a crucial role in determining the
particle spectrum shown in our next analysis, are insignificant in
the range $E\la E_1$. Therefore the particle spectrum produced by
the bubble should be $\sim E^{-\nu}$ for $E<10^{15}$ eV, where
$2> \nu >1$. As discussed in Section~\ref{sect:snr}, SNRs are
the major contributors for CRs with energies $E\la 10^{15}$ eV,
the exact particle spectrum generated from the bubble is
unimportant in the energy range of $E\la 10^{15}$ eV.

In the case of $\psi\ll 1$ or $l_D\gg l_{\rm sh}$,
inside the region of supersonic turbulence the acceleration regime
shifts to a pure stochastic acceleration by the supersonic
turbulence.
We extend the equation derived by \citet{byk90} for CR acceleration in supersonic turbulence
in a stationary state and axisymmetric geometry to include spatial dependent
diffusion coefficient and external source,
\begin{equation}
\frac{\partial}{\partial z}\left(D(\rho,p)\frac{\partial f}{\partial z}\right)
+\frac{1}{\rho}\frac{\partial}{\partial \rho}
\left(D(\rho,p)\rho\frac{\partial f}{\partial\rho}\right)
+\frac{1}{p^2}\frac{\partial}{\partial p}\left(\kappa(\rho,p)
p^2\frac{\partial f}{\partial p}\right)=-Q(\rho,z,p)\,, \label{dif_bubble}
\end{equation}
where $\rho$ and $z$ are the cylindrical spatial coordinates and $p$
is the particle momentum. $D(\rho,p)$ is the spatial diffusion
coefficient and $\kappa(\rho,p)$ is the momentum diffusion coefficient.
Their spatial and momentum dependence in our model is described below.
$Q(\rho,z,p)$ is the possible CR source which will be useful in our numerical
example in Section~\ref{sect:models}.

As we mentioned above, it is reasonable to assume that
CRs with energies $E\la 10^{15}$ eV are supplied by SNRs.
Therefore, in this section,
we concentrate on the analysis of the acceleration of CRs
in the energy range $E\ga E_1$ in the bubble by supersonic turbulence.
In Section~\ref{sect:models}, we will treat the case with SNRs and bubble
and deal with energy from less than $10^{12}$ eV to larger than $10^{18}$ eV.

We set the boundary condition of the distribution function at the Galactic halo
outer boundary
\begin{equation}\label{bubbleboundary}
f|_\Sigma =0\,,\quad{\rm at}\quad\rho=\rho_G\quad{\rm and}\quad z=\pm H\,.
\end{equation}
Proton acceleration in the bubble depends sensitively on the
acceleration parameters and structure of the bubble. In the
following we present a detailed analysis. We model the bubble
region as a cylinder extending above and below the Galactic plane
from $z=0$ to $z=\pm H $ with a radius $\rho=\rho_B$ and assuming
there is no CR source inside the bubble (i.e., $Q=0$).
The diffusion coefficients inside and outside the bubble are supposed
to be different
\begin{eqnarray}
&&D(\rho)=D_B\,\theta(\rho_B-\rho)+D_G\,\theta(\rho-\rho_B)\,, \label{spatialdiff} \\
&&\kappa(\rho,p)
=\kappa_B\,p^2\theta(\rho_B-\rho)\,, \label{diffusioncoeff}
\end{eqnarray}
where $D_B\sim cl_{\rm sh}/3$ is the coefficient inside the bubble as a result of
interactions with a supersonic turbulence and $D_G$ is the average
diffusion coefficient in the Galaxy, e.g., defined in \citet[][chap. 3, Section 6]{ber90}.
The momentum diffusion coefficient is $\kappa_B\sim u^2/D_B$.
The momentum dependence of $f$ is represented by a power-law
function, $f(p)\propto p^{-\gamma}$, where $\gamma$ should be
determined from Equation~(\ref{dif_bubble}).

To understand the dependence of $\gamma$ on other parameters,
we make two simplifications of Equation~(\ref{dif_bubble}),
which do not affect the value of $\gamma$ significantly.
First, for $H<\rho_G$,
as expected from \citet{strong},
particles that escape through the radial boundaries at $\rho=\rho_G$ are insignificant
\citep[see][chap. 3, Section 3]{ber90}, and we can shift the halo boundary to infinity,
i.e., $\rho_G=\infty$. Second, we model the axisymmetric geometry of
the problem as planar geometry
(i.e., we assume $\partial f/\partial\rho\gg f/\rho$).
We go back to the axisymmetric geometry afterward.

As in \citet{dog72} and \citet{dog74} we search for solutions to
Equation~(\ref{dif_bubble}) by the method of separation variables,
$f=R(\rho)Z(z)p^{-\gamma}$. The solution for $Z(z)$ has a very
simple form
\begin{equation}
Z_n(z)=\cos(k_n z/H)\,,
\end{equation}
where $k_n=\pi(n+1/2)$. We should point out that to make sure $f$ is non-negative
we must take $n=0$ for physical solutions.

To illustrate ideas, we consider the case $D_G=D_B$ and approximate the axisymmetric
geometry as planar (i.e., $d^2 R/d\rho^2\gg 1/\rho\,d R/d\rho$).
Using the dimensionless variable $\varrho=\rho/H$,
Equation~(\ref{dif_bubble}) can be simplified as
\begin{equation}
{d^2 R\over d\varrho^2}-
\left[k_n^2+\gamma(3-\gamma){\kappa_B\theta(\varrho_B-\varrho)H^2\over D_B}\right]R=0\,,
\label{dif_r}
\end{equation}
which has the exact form of the Schr\"{o}dinger equation for a rectangular
potential well:
\begin{equation}
{d^2\Psi\over d\varrho^2}
+{2m\over\hbar^2}
\left({\cal E}-U_0\right)\Psi=0\,,\quad{\rm for}\quad 0<\rho<\rho_B
\end{equation}
and
\begin{equation}
{d^2\Psi\over d\varrho^2}
+{2m\over\hbar^2}{\cal E}\Psi=0\,,\quad{\rm for}\quad \rho>\rho_B
\end{equation}
when we define ${\cal E}$ and $U_0$ as
\begin{equation}
{\cal E}=-\,{\hbar^2k_n^2\over 2m}\,\quad{\rm and}\quad
U_0={\hbar^2\over 2m}\gamma(3-\gamma){\kappa_B H^2\over D_B}\,.
\end{equation}
The solution of this equation is well-known \citep[e.g.,][ chap. 3, Section 22]{landau}:
\begin{eqnarray}
\Psi(\varrho)=&C_1\exp\left(k_n\varrho\right)\,,
&{\rm for}\ \varrho<0\,,\,{\rm where}\
k_n=\sqrt{-2m{\cal E}}/\hbar \nonumber \\
\Psi(\varrho)=&C_2\exp\left(-k_n\varrho\right)\,,
&{\rm for}\ \varrho>\varrho_B\,, \\
\Psi(\varrho)=&C\sin\left(\varsigma\varrho+\delta \right)\,,
&{\rm for}\ 0<\varrho<\varrho_B\,,\,{\rm where}\
\varsigma=\sqrt{2m({\cal E}-U_0)}/\hbar \nonumber
\end{eqnarray}
From the continuity of  the logarithmic derivative of $\Psi$ at
the well boundaries we have the condition
\begin{equation}
\arcsin\left(\frac{\hbar k_n}{\sqrt{-2mU_0}}\right)=\frac{(j\pi-k_n\rho_B)}{2}
\end{equation}
and
\begin{equation}
\cos\xi=\pm\chi\xi\,,\quad{\rm for\ odd}\ j\,, \label{cos}
\end{equation}
\begin{equation}
\sin\xi=\pm\chi\xi\,,\quad{\rm for\ even}\ j\,, \label{sin}
\end{equation}
where
\begin{equation}
\xi=\frac{k_n\rho_B}{2}\,\quad{\rm and}\quad
\chi=\frac{\hbar}{\rho_B}\sqrt{\frac{-2}{mU_0}}\,.
\end{equation}

For strong acceleration, $u^2 H^2/D_B^2\gg 1$, Equation~(\ref{cos}) and
(\ref{sin}) determine a finite number of levels. From these
equations, the first term of the series gives
\begin{equation}
\gamma=\frac{3}{2}+\sqrt{\frac{9}{4}+\frac{\pi^2D_B}{\rho_B^2\kappa_B}}\simeq 3\,.
\end{equation}

For weak acceleration, $u^2 H^2/D_B^2\ll 1$, we can use the shallow well
solution presented in \citet{landau}. In this case there is
only one level at $E_0\simeq U_0$ that gives
\begin{equation}
\gamma\simeq\frac{3}{2}+\sqrt{\frac{9}{4}+\frac{\pi^2D_B}{H^2\kappa_B}}\gg 1\,.
\label{g_est}
\end{equation}

A rough estimate of the power of CR production by the bubble
can be done in the same way as presented in \citet[][chap. 1, Section 4]{ber90} for GeV CRs.
The energy density of CRs at $E=3\times 10^{15}$ eV is
$n_{\rm CR}\simeq 6.7\times 10^{-17}$ erg cm$^{-3}$
\citep[see][]{kumiko}. Then the power required for the bubble to
produce the knee at the Earth is
\begin{equation}
W_B\sim \frac{cn_{\rm CR}M_H}{x}
\end{equation}
where $M_H$ is the total mass of hydrogen in our Galaxy, which is about $10^{43}$ g,
and $x$ is an extrapolation of energies $>10^{15}$ eV of the CR grammage derived from
the chemical composition by \citet{jones01} up to energies of about several hundred GeV,
$x(E)\sim 11.8\times (4.9\ {\rm GeV}/E)^{0.54}$ g cm$^{-2}$. Then we obtain the required power
of CR sources at the knee energy range $W_B\sim 2\times 10^{39}$ erg s$^{-1}$,
which can easily be supplied by star capture processes.

More accurate values of $\gamma$ can be derived from
numerical calculation of the axisymmetric case:
\begin{equation}
{1\over\varrho}{\! d\over d\varrho}
\left(\varrho{d R_i\over d\varrho}\right)-
\left[k_n^2+\gamma(3-\gamma){\kappa_{Bi}H^2\over D_B}\right]R_i=0\,.
\label{dif_rr}
\end{equation}
Here the index $i = 1,2$ denotes the regions inside ($\varrho<\varrho_B$) and outside
($\varrho\geq\varrho_B$) the bubble, respectively.
Note that $\kappa_{B1}=\kappa_B$ and $\kappa_{B2}=0$ (see Equation~(\ref{diffusioncoeff})).
A solution inside and outside the bubble is searched as
series of the Bessel functions ($J_\nu$, inside the bubble) and the McDonald
functions ($K_\nu$, outside the bubble).

\begin{figure}[t]
\centering
\epsscale{.80}
\plotone{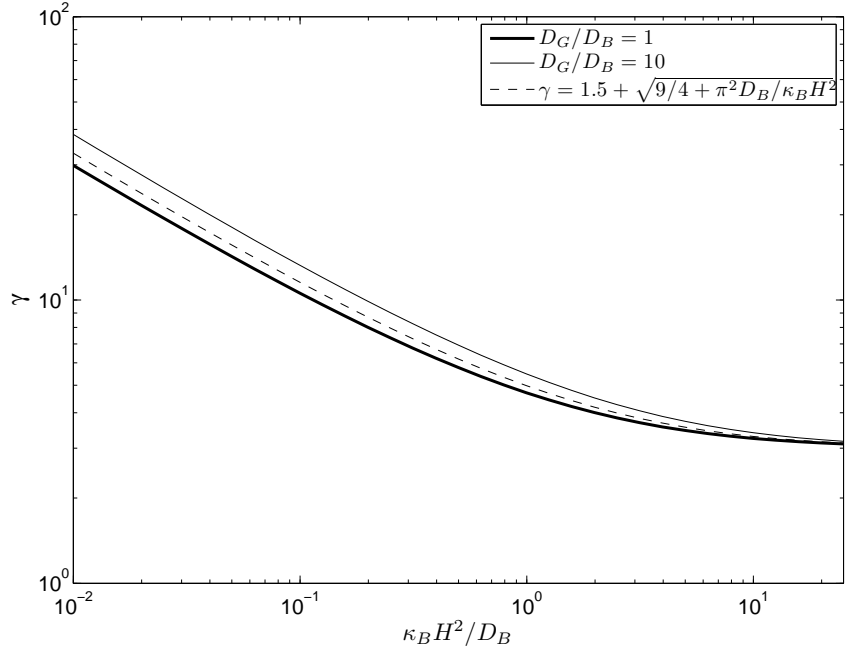}
\caption{Spectral index for different acceleration rates.
}
\label{fig:sindex}
\end{figure}

\begin{figure}[t]
\centering
\epsscale{.80}
\plotone{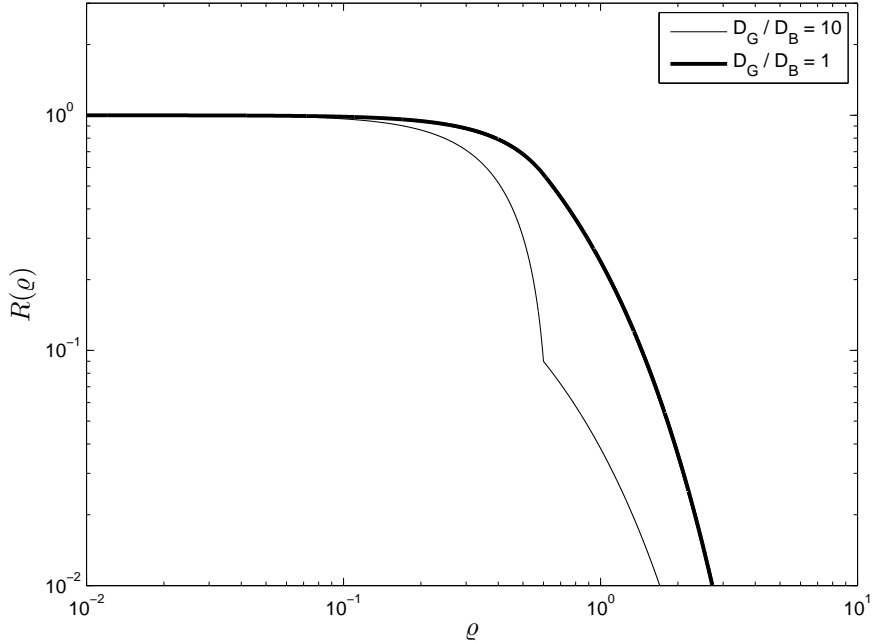}
\caption{Spatial distribution of protons in the central region of the Galaxy as a
function of $\varrho$.
Here $\varrho$ is normalized to $H$ (the height of the halo), and in this example
the bubble boundary is taken as $\varrho_B=0.6$.
}
\label{fig:spat}
\end{figure}

The boundary conditions at the bubble radius,
$\varrho = \varrho_B$ are
\begin{equation}\label{cont_rel}
R_1(\varrho_B) = R_2(\varrho_B)\,\quad{\rm and}\quad
D_B\left.{d R_1\over d\varrho}\right|_{\varrho_B}
= D_G\left.{d R_2\over d\varrho}\right|_{\varrho_B}\,.
\end{equation}
These relations
can be satisfied if
\begin{equation}
\alpha_1 = k_n^2 + \gamma(3-\gamma)\kappa_B H^2/D_B < 0,
\end{equation}
and the above requirement implies $\gamma >3$.
For $n=0$ we have
\begin{equation}
R_1(\varrho) = C_1 J_0(\sqrt{-\alpha_1}\varrho)\,\quad{\rm and}\quad
R_2(\varrho) = C_2 K_0(\pi\varrho/2)\,,
\end{equation}
or
\begin{equation}
\frac{D_B\sqrt{-\alpha_1}J_1(\sqrt{-\alpha_1}
\varrho_B)}{J_0(\sqrt{-\alpha_1 } \varrho_B)} = \frac{\pi D_G
K_1(\pi\varrho_B/2)}{2K_0(\pi\varrho_B/2)}\,.
\end{equation}
Numerical results of different ratios of
$D_G/D_B$ as a function of the ratio between the times of particle
escape from the Galaxy $H^2/D_B$ and the time of Fermi
acceleration $1/\kappa_B \sim D_B/u^2$ are shown in Figure~\ref{fig:sindex}. The
dashed curve shows the approximate estimate given by
Equation~(\ref{g_est}), which indicates that Equation~(\ref{g_est}) is
a good approximation and we will use it in the next section of
data fitting for simplicity.

The spatial distribution of accelerated protons for different
ratios $D_G/D_B$ is shown in Figure~\ref{fig:spat}.
We can see that these two distributions do not have a qualitative difference even
if the ratios differ by a factor of 10.

\section{BUBBLE CONTRIBUTION TO CRs AT ENERGY LARGER THAN $10^{15}$ eV}\label{sect:beyondknee}

In this section, we use the general models developed in the last
section to fit the CR spectrum for $E>10^{15}$ eV but we ignore any possible
spectral modulation effect as a result of propagation from the bubble to
the Earth. We have shown that the charged particles in the bubble
can be described by a  power-law spectrum,
\begin{equation}
\frac{d\dot{N}}{dE}=\frac{\dot{N}_{0}}{E_1}\left({E\over E_1}\right)^{-\nu}\,,
\label{spectrum}
\end{equation}
where $E_1$ is given by Equation~(\ref{e1})
and $\nu=\gamma -2$, which can be expressed as
\begin{equation}\label{spectralindex}
\nu=-\frac{1}{2}+\sqrt{\frac{9}{4}+\frac{\pi^2D_B^2}{u^2H^2}}\,,
\end{equation}
for $E_1<E<E_2$ where $E_2$ is the high-energy cutoff and will be discussed below.
We have used the momentum diffusion coefficient
$\kappa_B=u^2/D_B$. Here $\dot{N}_{0}$ is a
normalization constant which is chosen to fit the observed CR
spectrum by assuming that the CR spectrum with energy
$E>10^{15}$ eV is entirely contributed by particles from the
bubble. $E_2$ can be estimated by Equation~(\ref{emax}), which gives
\begin{equation}\label{e2min}
(E_2)_{\rm min} \approx 3\times 10^{17}\left({B\over 5\, \mu{\rm G}}\right)
\left({T\over 10\,{\rm Myr}}\right)\left({u\over 10^8\,{\rm cm\,s}^{-1}}\right)^2
\ {\rm eV}\,,
\end{equation}
where $T$ is the time taken by a single shock propagating from the disk to the top of the bubble.
However, this estimate assumes that particles are only accelerated by a single shock.
As described in the previous section, the high-energy particles are accelerated by multiple shocks
in the bubble; in other words, particles can diffuse downward and continue to be accelerated
by younger shocks in the lower part of the bubble. Therefore $T\sim 10$ Myr is only a minimum
lifetime, which means that the above estimate can be considered
as the lower limit of $E_2$. Another possible way to restrict
the maximum value of $E_2$ is when the Larmor radius of the particles
($r_L=E/eB$) is larger than the radius of
the bubble ($H/2$),
\begin{equation}\label{e2max}
(E_2)_{\rm max}\approx 10^{19}\left({B\over 5\,\mu{\rm G}}\right)
\left({H\over 10\,{\rm kpc}}\right)\ {\rm eV}\,.
\end{equation}
In fact this estimate should be more appropriate for $E_2$ and
we will use it in our model fitting process.

\begin{figure}
\epsscale{.80}
\plotone{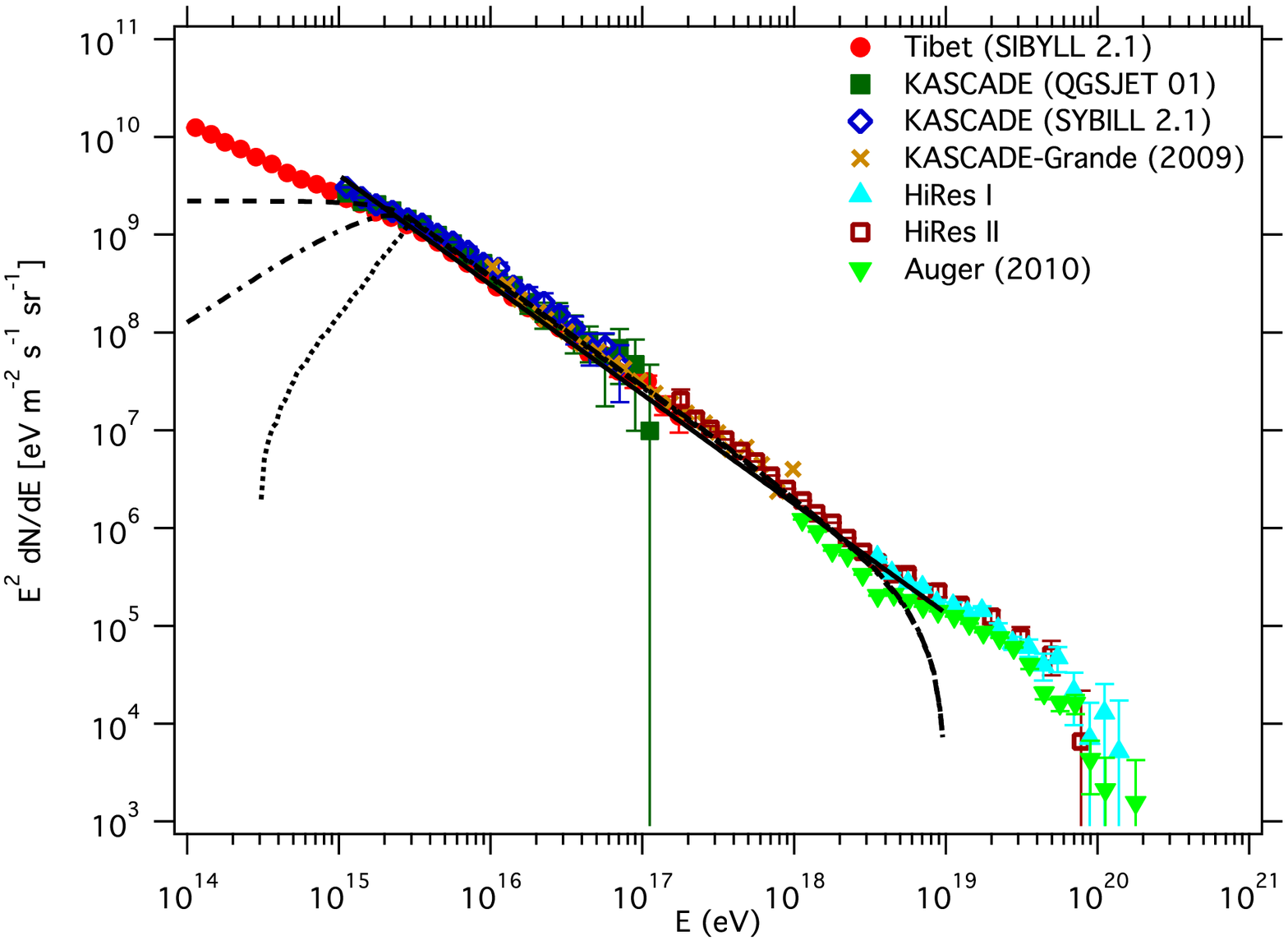}
\caption{Bubble contribution to the flux of CRs.
The data are summarized in \citet{kumiko}.
The spectrum has been multiplied by $E^2$ for clarity.
Experiments include
Tibet AS-$\gamma$
\citep{Amenomori}
KASCADE \citep{Kampert},
KASCADE-Grande \citep{Apel},
High Resolution Fly's Eye cosmic-ray detector-I
\citep[HiRes-I;][]{Abbasi2009},
HiRes-II \citep{Abbasi2008},
and Auger \citep{Abraham}.
The solid line shows the contribution from the bubble predicted
by Equation~(\ref{spectralindex}) for $E_1=10^{15}\,{\rm eV} < E < E_2=10^{19}\,{\rm eV}$.
The dotted line has the same spectral index as that of the solid line but
both $E_1$ and $E_2$ are assumed to obey a uniform distribution
(see the text for explanation). The dashed line and the dash-dotted line
have the same
spectral index for $E_1<E<E_2$ and
distribution of $E_1$ and $E_2$
as that of the dotted line except the spectral index changes to 1 and 2, respectively,
for $E < E_1$ (see the text for explanation).
}
\label{fig:flux}
\end{figure}

In our model particle spectrum there are four parameters:
the lower cutoff $E_1$ (Equation~(\ref{e1})), the upper cutoff $E_2$ (Equation~(\ref{e2max})),
the spectral index $\nu$ (Equation~(\ref{spectralindex})), and the normalization
${\dot N}_0$ (Equation~(\ref{spectrum})).
We consider that $l_{sh}=30$ pc, $B=5$ $\mu$G, $u=10^8$ cm s$^{-1}$ and $H=10$ kpc are very
reasonable mean values in the bubble, therefore in our model fitting we fix
$E_1=10^{15}$ eV and $E_2=10^{19}$ eV.
On the other hand, the conversion efficiency from the shock energy into particle energy
and the  mean diffusion coefficient $D_B$ in the bubble are the most uncertain parameters.
Therefore in our model fitting we treat these two parameters as fitting parameters.
The solid line in
Figure~\ref{fig:flux} indicates the best fit.
The best fit gives $\nu =3.12$, which corresponds to $D_B\sim 3\times 10^{30}$ cm$^2$ s$^{-1}$.
This is in agreement with the estimation $D_B\sim cl_{\rm sh}/3$ if we take the
average separation between shocks as $l_{\rm sh}\sim 100$ pc
(see Equation~(\ref{l_shock})).
This seemingly large diffusion coefficient is in fact at least an order of magnitude smaller
than the coefficient in the halo for $E > 5\times 10^{15}$ eV
\citep[e.g.,][ suggested $D_G\approx 2.0\times 10^{28}(E/4.9\,{\rm GeV})^{0.54}$ cm$^2$ s$^{-1}$]
{jones01}.

If the CRs in the energy range $10^{15}$ eV $<E< 10^{19}$ eV arriving at the Earth come from
the bubble, then the power provided by the bubble for CR in this energy range
is given by
$\dot{W}_{\rm CR} \approx \int_{10^{15}{\rm eV}}^{10^{19}{\rm eV}}4\pi R^2 F_{\rm CR}(E)dE
\sim 3\times 10^{39}(R/10\,{\rm kpc})^2$ erg s$^{-1}$,
where $F_{\rm CR}(E)$ is the observed CR energy flux and $R$ is the mean distance to the bubble.
We find that the conversion efficient from shock power,
$\dot{W}\sim W/\tau_{\rm cap}\sim 3 \times 10^{40}$ erg s$^{-1}$, is about 10\%,
which is consistent with recent estimation by using the {\it Fermi}-LAT data \citep{Abdo2010}.
Realistically, particles could escape from the bubble through various locations
of the bubble's surface, where the local strength of the magnetic field may be different.
In addition, we have pointed out that there are two possible ways to estimate the value of
$E_2$ (Equations~(\ref{e2min}) and (\ref{e2max})).
Therefore it is likely that $E_1$ and $E_2$
should have some distribution coming out from the bubble.
In the dotted line in Figure~\ref{fig:flux}
we keep the spectral index as that of the solid line but assume that
$E_1$ and $E_2$ have a uniform distribution between
$2\times 10^{14}$ eV $<E_1< 2\times 10^{15}$ eV (which corresponds
to 1 $\mu$G $<B<$ 10 $\mu$G) and $(E_2)_{\rm min}<E_2<(E_2)_{\rm max}$, respectively.
We can see that the model curve starts to drop around $3\times 10^{18}$ eV and matches the
data better. Furthermore, we have argued that the
injected spectrum from the shock is $E^{-2}$ for $E<E_1$ and slowly becomes harder for
$E^{-1}$ due to interactions with multiple shocks after a sufficiently long time.
However, it is unclear if this spectral hardening process can be completed within the finite
lifetime of the particles in the bubble (see Section~\ref{sect:bubbleshock}
for a more detailed discussion).
Therefore we cannot predict the exact spectral index for $E<E_1$.
For reference purposes, in Figure~\ref{fig:flux} we show these two possibilities:
the dashed line for $E^{-2}$ and the dash-dotted line for $E^{-1}$, respectively.
Both $E_1$ and $E_2$ of these two lines also assume uniform distributions.

\section{MODEL FOR HIGH-ENERGY CRs WITHIN AND BEYOND THE KNEE}\label{sect:models}

The observed high-energy CR spectrum is a broken power law.
The energy spectral index for CR energy smaller than the knee
(around $3\times 10^{15}$ eV) is 2.7.
The index increases to 3.1 for larger energy (until around $10^{19}$ eV where
the spectrum becomes harder again).
The general argument is that the sources (i.e., acceleration mechanisms and/or sites)
responsible for the energy range within and beyond the knee should be different.
The intriguing fact is that the two power laws match quite well at the knee.
This coincident problem is difficult to solve if the two sources are totally unrelated.

CRs within (i.e., energy less than) the knee are generally attributed to
SNRs in our Galaxy (see Section~\ref{sect:snr}).
In the previous two sections, we alluded to the acceleration site of CRs beyond the knee
(i.e., energy larger than the knee) to the Fermi Bubble.
In Section~\ref{sect:bubbleshock}, we studied Equation~(\ref{dif_bubble}) without a source in
the bubble. The solution gave the characteristic spectrum of the system.
In reality we need a source or a seed population for the bubble.
We deem that we do not need a new seed population for bubble acceleration.
Instead we propose the following model:
some of the CRs produced by SNRs in the Galactic
disk are re-accelerated in the bubble to energy beyond the knee.
This model has the potential to solve the coincident
problem naturally, because the source of CRs beyond the knee
is seeded by the source within the knee.

To consolidate our idea, we work out a concrete numerical model.
Essentially, we solve the stationary state CR
transport equation~(\ref{dif_bubble}) in our Galaxy with
two Fermi Bubbles (one on each side of the Galactic plane).
We modeled our Galactic halo as a cylinder of
radius $\rho_G=20$ kpc, and the top and bottom at $\pm 10$ kpc
from the mid-plane. Each Fermi Bubble is also a cylinder of the
same height $\pm 10$ kpc, but with a radius $\rho_B=3$ kpc.

The Fermi Bubbles are filled with shocks as described
in Section~\ref{sect:structure} (see Figure~\ref{fig:shape}).
The spatial diffusion coefficient are different inside and outside the bubble as
described by Equation~(\ref{spatialdiff}).
Due to the very turbulent condition inside the bubble, we consider a constant spatial diffusion
coefficient and adopt $D_B=2.08\times 10^{30}$ cm$^2$ s$^{-1}$
(cf. estimate value by the fitting process in Section~\ref{sect:beyondknee}).
Outside the bubble, we take into account the energy (or momentum)
dependence of the spatial diffusion coefficient and adopt
$D_G=D_0(pc/4\,{\rm GeV})^{0.6}$, $D_0=6.2\times 10^{28}$ cm$^2$ s$^{-1}$
\citep[cf.][]{jones01}.

According to the analysis in Section~\ref{sect:bubbleshock}, the acceleration of energetic
particles in the bubble is facilitated by stochastic acceleration
(second-order Fermi acceleration).
Assuming that there is little or no stochastic acceleration outside the bubble, we
model the momentum diffusion coefficient as a step function as in Equation~(\ref{diffusioncoeff})
and adopt $\kappa_B H^2/D_B=1.9$ (i.e., $\kappa_B=4.4\times 10^{-15}$ s$^{-1}$ or
the corresponding acceleration time scale is $7.6$ Myr).

The Galactic disk contains SNRs.
We adopt the distribution suggested by \citet{stecker}
and modified it with a Gaussian thickness profile
\begin{equation}\label{SNRdist}
Q_{\rm SNR}(\rho,z)\propto \left({\rho\over R_\odot}\right)^{1.2}
\exp\left(-\,{3.22\rho\over R_\odot}\right)\exp\left(-{z^2\over h^2}\right)\,,
\end{equation}
where $\rho$ is the galactocentric radius and $z$ is the distance perpendicular to the
mid-plane.
Here we take $h=100$ pc, $R_\odot=8$ kpc.
We adopt the idea that SNRs inject energetic particles in the form of a power law with
a high-energy cutoff at $p_{\rm max} c\approx 3\times 10^{15}$ eV. Therefore, together
with the SNR distribution (Equation~(\ref{SNRdist})), the source function is
\begin{equation}
Q(\rho,z,p)=Q_0 \left({p\over p_{\rm max}}\right)^{-\mu}\exp\left(-\,{p\over p_{\rm max}}\right)
\left({\rho\over R_\odot}\right)^{1.2}
\exp\left(-\,{3.22\rho\over R_\odot}\right)\exp\left(-{z^2\over h^2}\right)\,.
\end{equation}

As mentioned in Section~\ref{sect:snr}, we take the SNR injection
spectrum to be $\mu=4.35$ \citep[see also][]{biermann}.
The normalization $Q_0$ is obtained by fitting the simulation result to the observed spectrum
and the value is
$1.5\times 10^{14}$ particles s$^{-1}$ kpc$^{-3}$ (GeV/c)$^{-3}$.
Integrating $Q(\rho,z,p)$ over the Galaxy and momentum (from 1 GeV c$^{-1}$ to $p_{\rm max}$)
gives the total luminosity of CRs $4\times 10^{40}$ erg s$^{-1}$,
which is consistent with the value in the  literature \citep[e.g.,][chap. 1, Section 4]{ber90}.

Finally, the appropriate boundary conditions for the momentum
coordinate are
\begin{equation}
\left.{p\over f}{\partial f\over\partial p}\right|_{p=p_{\rm low}}= -4.7\,,\quad
f|_{p=p_{\rm up}} = 0\,,
\end{equation}
where the energy of the lower momentum boundary is $p_{\rm low}c = 10^{12}$ eV, and
the upper momentum boundary is $p_{\rm up}c = 3\times 10^{18}$ eV.
The condition at the lower momentum ensures that the spectral index matches that of
low-energy CRs (say $E<10^{12}$ eV).

The spatial boundary conditions are
\begin{equation}
\left.{\partial f\over\partial\rho}\right|_{\rho=0} = 0\,,\quad f|_{\rho=\rho_G} = 0\,,
\end{equation}
where the radius of the Galactic disk was taken to be $\rho_G = 20$ kpc.
\begin{equation}
\left.{\partial f\over\partial z}\right|_{z=0} = 0\,,\quad f|_{z=\pm H} = 0\,,
\end{equation}
where the height of the halo $H= 10$ kpc.

\begin{figure}[t]
\centering
\epsscale{.80}
\plotone{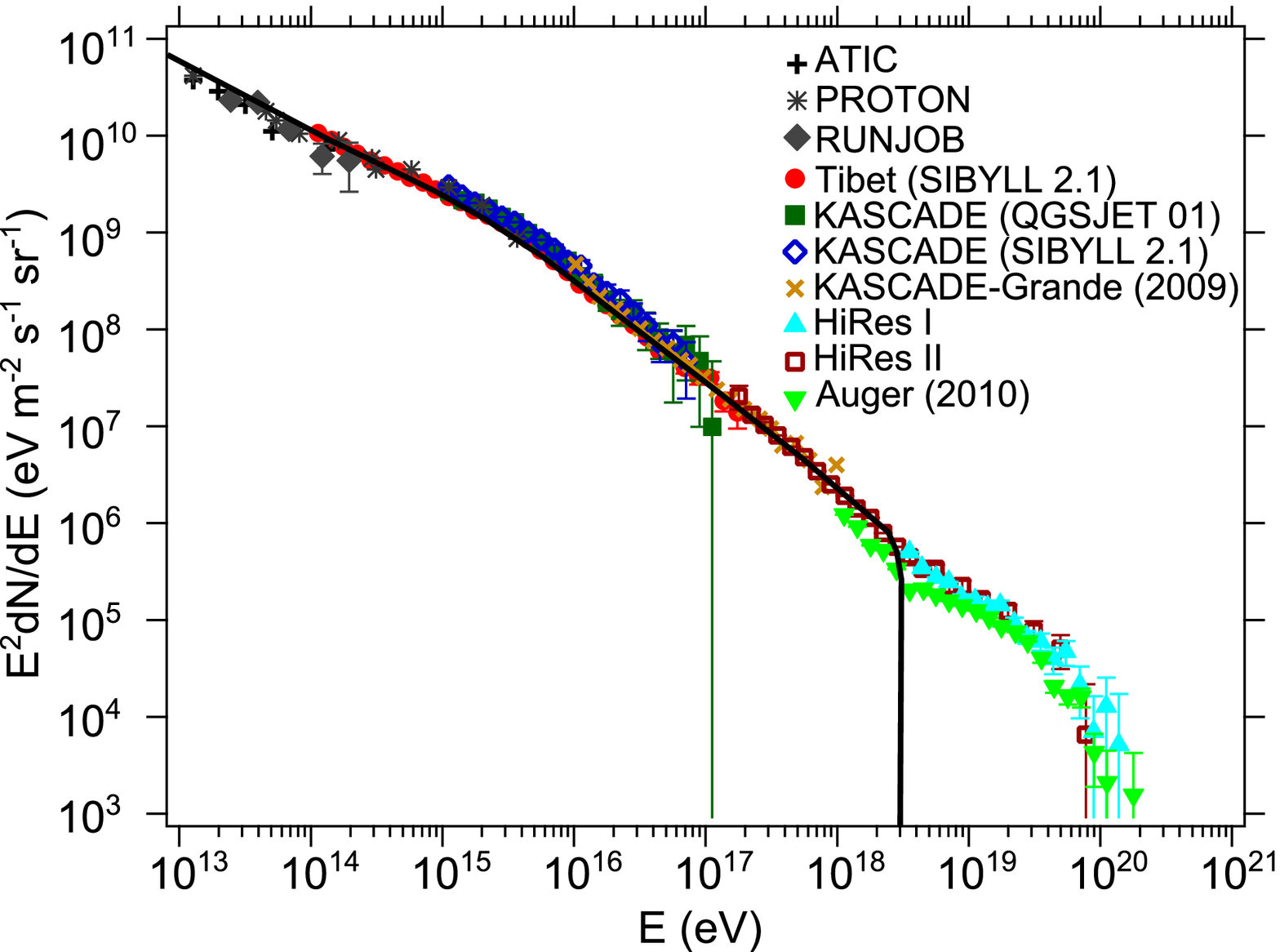}
\caption{CR spectrum at the Earth as a combination of the SNR contribution (in the Galactic disk)
and the stochastic acceleration in the Fermi Bubbles.
In addition to data from experiments presented in Figure~\ref{fig:flux}, we added experiments
for lower energies:
ATIC \citep{Ahn},
Proton \citep{Grigorov} and
RUNJOB \citep{Apanasenko}.
The black solid line is the spectrum from our numerical model.
In this model,
$D_B=2.08\times 10^{30}$ cm$^2$ s$^{-1}$ inside the bubbles
and $D_G=6.2\times 10^{28} (pc/4\,{\rm GeV})^{0.6}$ cm$^2$ s$^{-1}$ outside,
$\kappa_B H^2/D_B=1.9$, and the injection spectrum from SNR $\mu=4.35$.
}
\label{fig:dif+fermi2}
\end{figure}

\begin{figure}[t]
\centering
\epsscale{.80}
\plotone{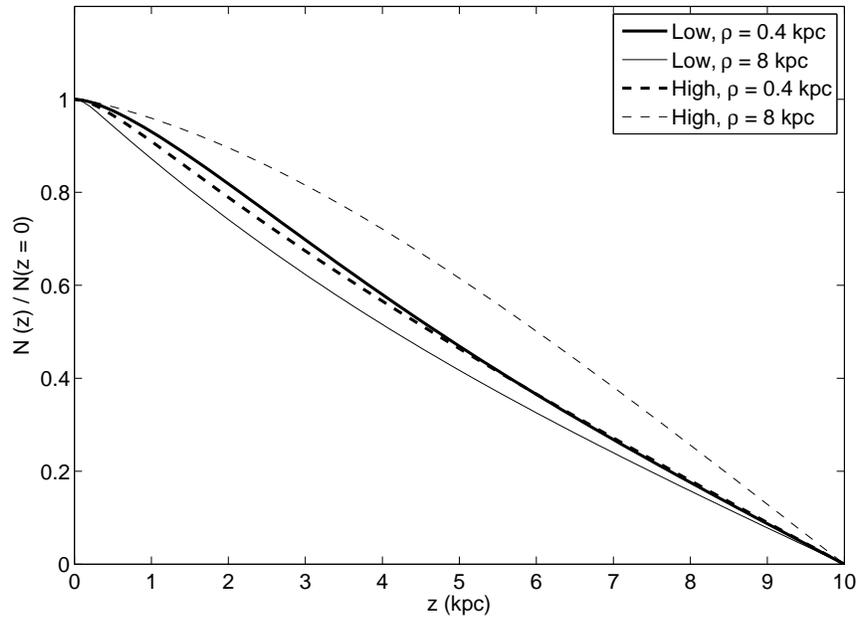}
\caption{
Vertical distribution of the number density of CR at two galactocentric radii,
0.4 kpc (thick lines) and 8 kpc (thin lines), in two energy ranges,
``Low'' for $1\times 10^{13}\sim 3\times 10^{15}$ eV (solid lines) and
``High'' for $3\times 10^{15}\sim 3\times 10^{18}$ eV (dotted lines).
}
\label{fig:dstr_z}
\end{figure}

\begin{figure}[t]
\centering
\epsscale{.80}
\plotone{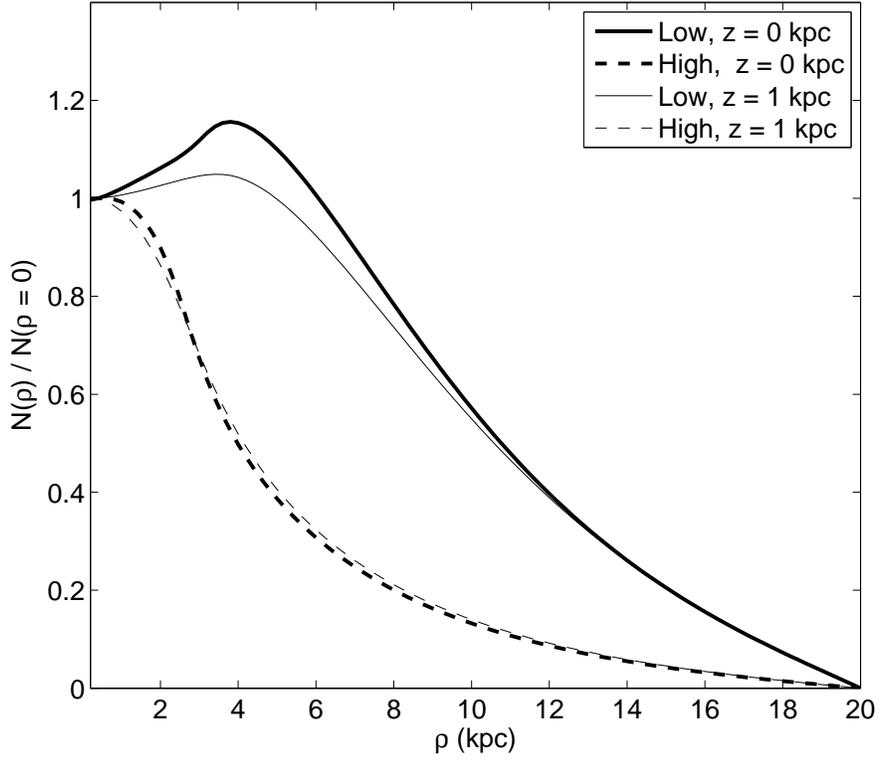}
\caption{
Radial distribution of number density of CR at the Galactic plane
(thick lines) and 1 kpc above the plane (thin lines), in two energy ranges,
``Low'' for $1\times 10^{13}\sim 3\times 10^{15}$ eV (solid lines) and
``High'' for $3\times 10^{15}\sim 3\times 10^{18}$ eV (dotted lines).
}
\label{fig:dstr_r_new_lin}
\end{figure}

\begin{figure}[t]
\centering
\epsscale{.80}
\plotone{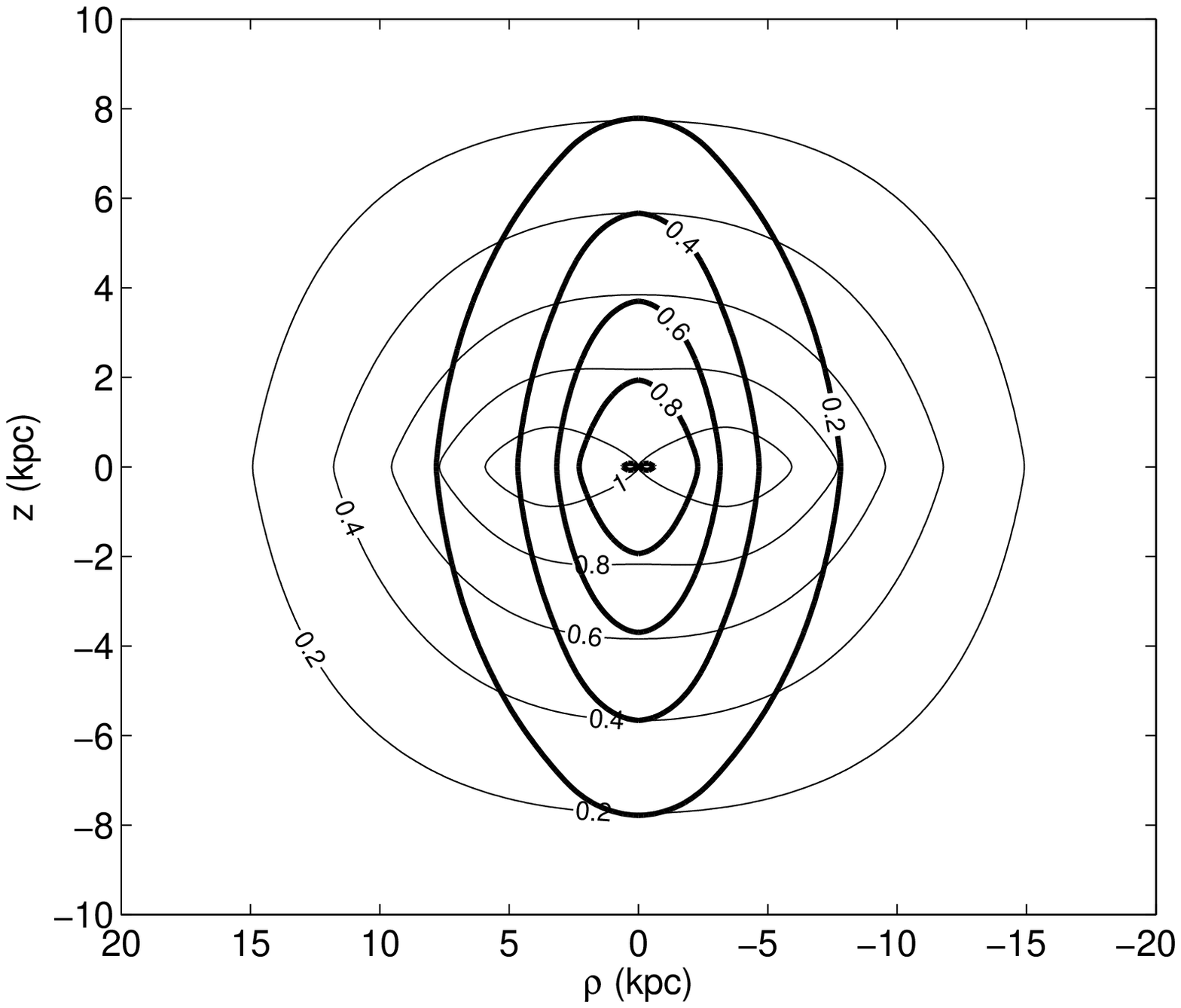}
\caption{
Contour of the relative number density distribution of the re-accelerated CRs
($E>3\times 10^{15}$ eV) in the halo (thick lines).
The seed for the re-accelerated CRs comes from SNRs in the Galactic plane (thin lines).
}
\label{fig:isolines_new}
\end{figure}

The spectrum evaluated at Earth's position is the solid line shown
in Figure~\ref{fig:dif+fermi2}.
The model fits the data reasonably well and it is not coincident that the
spectra join smoothly at the knee.
For reference, in Figures~\ref{fig:dstr_z} and \ref{fig:dstr_r_new_lin} we show the spatial
distribution of CRs from our simulation.
The labels ``Low'' and ``High'' refer to the number density of particles in the energy range
$1\times 10^{13}\sim 3\times 10^{15}$ eV and $3\times 10^{15}\sim 3\times 10^{18}$ eV,
respectively.
Figure~\ref{fig:isolines_new} is a contour plot of the number density distribution of
re-accelerated CRs ($E>3\times 10^{15}$ eV) (thick lines).
In the figure we also plot the distribution of seed particles from SNRs (thin lines).
We point out that the spatial distributions of seed and re-accelerated CRs in the disk are
quite different.
In principle, CR distribution can be derived from gamma-ray data.
For instance, \citet{Breitschwerdt2002} used the gradient of gamma-ray emissivity in the disk
derived from the EGRET data for the analysis of CR propagation in the Galaxy.
If the diffuse gamma-ray data at $E>10^{15}$ eV were available, the gradient test would be
a nice tool to investigate possible proton sources in this energy range and might lend support
to our model.

\section{DISCUSSIONS AND CONCLUDING REMARKS}\label{sect:summary}

We have summarized our current understanding of the origin of CRs.
It is generally believed that most CR power can be provided by SNRs.
However the CRs with energies $E>10^{15}$ eV are quite difficult to achieve in SNe
due to the limited acceleration time and energy content in SN shocks.
On the other hand, we argue that shocks in the Fermi Bubble produced by stellar capture
events can have a much longer lifetime $>10^7$ yr and larger energy content
$\sim 3\times 10^{52}$ erg, which allow them to produce CRs with energies $E>10^{15}$ eV.
If processes of CRs which escape from the Galaxy are taken into account,
the predicted CR spectrum contributed by the bubble is $E^{-\nu}$, where
$\nu=-\frac{1}{2}+\sqrt{\frac{9}{4}+10\frac{(D/3\times
10^{30}\,{\rm cm}^2{\rm s}^{-1})^2}{(u/10^8\,{\rm cm\,s}^{-1})^2(H/10\,{\rm kpc})^2}}\sim 3$
for $10^{15}$ eV $< E < 10^{19}$ eV.
However, it is very difficult to predict the exact value of the spectral index $\nu$ due to
the poorly measured value of the diffusion coefficient in the bubble.
So we fit the observed CR spectrum between $10^{15}$ eV
and $\sim 10^{19}$ eV and find that the diffusion coefficient is about
$3\times 10^{30}$ cm$^2$ s$^{-1}$.
By matching with the observed flux in the knee region we find that the conversion
efficiency from shocks in the bubble to
CRs is about 10\%, which is quite consistent with numerical simulations.
Other input parameters in this model, such as the capture time scale
$\tau_{\rm cap}\sim 3 \times 10^4$ yr, the mean rate of energy release
$\dot{W}\sim 3 \times 10^{40}$ erg s$^{-1}$ per capture and the injected
wind speed $u\sim 10^8$ cm s$^{-1}$, have been estimated and used in other
observed phenomena in GC \citep[e.g.,][]{cheng1,cheng2,dog_pasj1,dog_pasj,dog_aa,dog_pasj2011}.

We put forth the idea that the Fermi Bubble acts as the re-acceleration site for the
CRs produced by SNRs in the Galactic disk.
The re-accelerated particles form the part of the observed spectrum that is beyond the
knee (about $3\times 10^{15}$ eV).
The part within the knee is formed by the CRs produced by the SNRs.
We demonstrated this idea by a numerical model.
We solve the stationary transport equation in our Galaxy with two Fermi Bubbles
(see Section~\ref{sect:structure} for our model of the bubbles).
The re-acceleration process in the bubble is facilitated by stochastic acceleration.
Our model simulated the observed spectrum nicely.
Therefore we consider that this model provides a natural explanation of the flux,
spectral index and matching at the knee of CRs in this energy range.
In a related issue, \citet{Mertsch2011} showed that the gamma ray from the Bubbles can be
produced by stochastic acceleration of electron throughout the Bubbles.

As described in Section~\ref{sect:structure} there are many shocks propagating in the
Fermi Bubbles. After being re-accelerated inside the bubbles by the multiple shocks, protons
(and nuclei) escape the bubbles.
The lifetime of protons by pp collision is of the order of $10^9\sim 10^{10}$ yr
\citep[e.g.,][]{crock}.
The diffusion time for protons at these energies to escape the Galaxy is of the order of
$10^7$ yr.
Thus after leaving the bubbles, protons (and nuclei as well) diffuse throughout the whole Galaxy
(including the Earth) without any attenuation of energy.
Not only can protons (and nuclei) be accelerated by the multiple shocks, but so can electrons.
Nevertheless, energetic electrons lose energy efficiently.
The lifetime of electrons can be estimated by $\tau_e=1/\beta_e E$ (e.g., inverse Compton
and synchrotron).
Taking $\beta_e=3\times 10^{-25}$ eV$^{-1}$ s$^{-1}$, the lifetime of electrons of energy
$0.1 \sim 1$ TeV is $1\sim 0.1$ Myr.
Once they leave the acceleration site (i.e., Fermi Bubbles), they lose most of their
energy within a short distance (less than 1 kpc).
Hence the electrons are mostly confined in the bubbles.

\citet{meng} suggested that the electron spectrum must be $E^{-\alpha}$ for $E<$ TeV and
the spectral index $\alpha \sim 2.4$-$2.8$. It is clear that the energy loss processes
for protons and electrons are very different, therefore protons and electrons in the bubble
can have different spectra.
In the CCDKI model we have assumed that the shocks can produce an injected electron spectrum
$\sim E^{-2}$, in which the spectrum is modified by processes of energy losses and escape.
In our subsequent work we will take into account the stochastic acceleration processes
(multiple-shock) to see how the electron spectrum is affected.
We will also further analyze the spatial spectral distribution of electrons.
Preliminary results can be found in \citet{chernyshov2011},
in which the spatial distribution of gamma=ray emission is reproduced nicely
by the multi-shock model.

\section*{Acknowledgements}

We are very grateful to A. M. Bykov, R. M. Crocker, A. D. Erlykin,
Y. Uchiyama and V. N. Zirakashvili for useful discussions,
and K. F. Sinkov who performed graphic and numerical
illustrations of the Kompaneets' solution presented in Section~\ref{sect:structure}.
D.O.C. and V.A.D.  are partly supported by the NSC-RFBR
Joint Research Project RP09N04 and 09-02-92000-HHC-a.
K.S.C. is supported by the GRF Grants of the Government of the Hong Kong SAR
under HKU 7011/10P.
C.M.K. is supported by the Taiwan National Science Council Grants
NSC 98-2923-M-008-01-MY3 and NSC 99-2112-M-008-015-MY3.
W.H.I. is supported by the Taiwan National
Science Council Grant NSC 97-2112-M-008-011-MY3 and Taiwan
Ministry of Education under the Aim for Top University Program
National Central University.


\end{document}